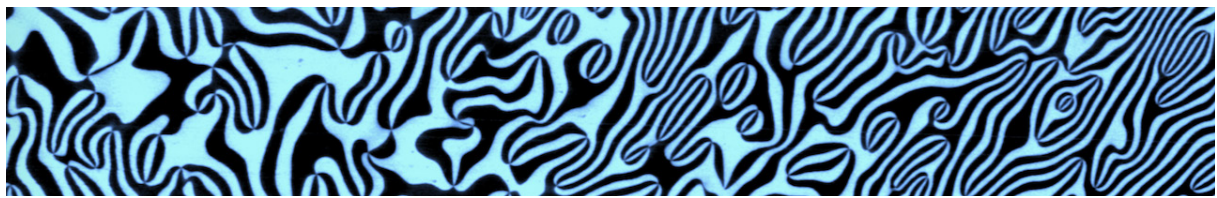

*Piezoresponse force microscopy image of the ferroelectric domain structure of multiferroic ErMnO$_3$. The dark and light regions correspond to opposite orientations of the electric dipoles. The vertical dimension is about 20 μm.*
*Image courtesy of Manfred Fiebig and Martin Lilienblum (ETH Zürich)*

# Fundamental Materials Research and the Course of Human Civilization

**Nicola Spaldin\***

| *Unless we change direction, we are likely to wind up where we are headed.* (Ancient Chinese proverb) |
| --- |

The field of Materials Science has at its core the science and engineering of "useful stuff": We materials scientists make materials for shelter, tools, transportation, information technologies, health, as well as recreation and the arts… all of the essentials and luxuries without which our lives would be unimaginable. As a result of this universal utility, the value of applied research and development in Materials Science is rarely called into question. In this issue of the "Bulletin VSH-AEU" concentrating on fundamental research, I focus on the necessity of basic materials research, and make the case that, because of its central role and unique relevance, fundamental research in Materials Science is a cornerstone of human progress.

From the Stone Age through the Bronze and Iron Ages to today's Silicon Age, every major advance in human civilization has been driven by a fundamental development in Materials Science. The transition from Hunter-Gatherer societies to the adoption of agriculture at the Neolithic revolution coincided with developments in processing techniques for natural materials such as stone so that tools for grinding and cutting could be produced. The discovery of material composites led to, for example, the attachment of stone blades to wooden handles with fibres or resins, providing additional leverage over hand-held tools. Metals were at first a rarity, and native nuggets of copper, silver and gold were used primarily as ornaments for ceremonial purposes. But the discovery of the smelting process to extract metals from mineral ores ushered in a new era, the Bronze Age, associated with the establishment of cities and the beginning of crafts and trade. It's hard to appreciate what a remarkable breakthrough in basic metallurgy the development of bronze represents: First, the smelting process requires temperatures above the melting point of the metal (~1000° C for copper) as well as a reducing atmosphere, conditions that were prob-ably first achieved accidentally in kilns designed for firing pottery. Second, the favourable properties of bronze result from the alloying of different elemental metals (in this case copper and tin), and the detailed fundamental physics of why alloying so profoundly improves the behaviour remains a topic of current research interest! But it's clear that without this basic materials discovery, which led to a complete change of direction in the evolution of human progress, the world today would be a vastly different place. Later, the adoption of iron – which can be hammered rather than cast – in the Iron Age drove radical changes in agriculture as well as of course weaponry leading to the establishment of countries and empires and coinciding with the beginning of written literature. Interestingly, many of these early scientific developments were disseminated through trade, or what we would now call international collaboration.

Iron continued to play a central role throughout history, culminating after around 4000 years of developments in metallurgy with the industrial revolution. But a profound change in direction was initiated at

\* ETH Zürich, Professur für Materialtheorie,
Wolfgang-Pauli-Strasse 27, HIT G 43.3, 8093 Zürich.

E-mail: *nicola.spaldin@mat.ethz.ch*

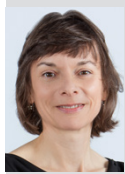

**Nicola Spaldin,** PhD FRS, is the Professor of Materials Theory at ETH Zürich. She was educated at Cambridge University (BA and MA in Natural Sciences) and UC Berkeley (PhD in Chemistry). Following postdoctoral work at Yale University in Applied Physics, she established her independent career at UC Santa Barbara as Assistant, Associate then Full Professor, before moving to ETH in 2011. She is a passionate science educator, director of her department's study program, and holder of the ETH Golden Owl Award for excellence in teaching. She developed the class of materials known as multiferroics, which combine and couple ferromagnetism and ferroelectricity. When not trying to make a room-temperature superconductor, she can be found playing her clarinet, or skiing or climbing in the Alps.

the end of the 19th century by a remarkable event: The discovery of the electron [1]. In his laboratory in Cambridge, J. J. Thomson was performing detailed fundamental experiments to try to figure out what the mysterious radiation emitted by negative metallic electrodes – so-called "cathode rays" – was made of. Thomson was able to show that the rays are made of particles that are negatively charged with a mass almost 2000 times lighter than a hydrogen atom. Soon it was recognized that Thomson's cathode ray particles are the same as those that carry current in wires and they became known as electrons. Thomson's discovery was a profound breakthrough in fundamental sub-atomic physics, for which he received the 1906 Nobel Prize in Physics *"in recognition of the great merits of his theoretical and experimental investigations on the conduction of electricity by gases"*. I doubt that he could have imagined, though, that his fundamental research would so dramatically change the course of human history – that he had made the first steps to ushering in a new age – by paving the way for the development of electronic devices.

In fact, the equipment that Thomson developed for his fundamental studies – glass tubes strong enough to withstand the pumping out of most of the air molecules – provided engineers with the design for the first electronic devices: Three-terminal devices called vacuum tubes or triodes which could be used as switches, for amplification, and to make simple logic circuits. Vacuum tubes offered the first glimpse of the tremendous capabilities that might be provided by electrical circuits. Since, however, they were bulky, sometimes unreliable and devoured energy, they also provided device physicists and materials scientists with a strong incentive to find a route to a more convenient three-terminal device. The resulting development of the semiconductor transistor, first in germanium and soon after in silicon because of its superior material properties, is certainly one of the most significant breakthroughs of the 20th Century. It is interesting to reflect on the enablers for this giant leap forward: The relevant properties of semiconductors are fundamentally quantum mechanical, and a prerequisite to the development of the transistor was the development of quantum mechanics and the subsequent decades of fundamental research on the quantum theory of solids. Equally important was the tremendous technological progress that had been made in the properties of silicon and germanium, particularly in producing materials of very high purity, which was motivated by the need for high-frequency radar receivers during the Second World War. And finally, the vision of the management (and indeed the shareholders) at Bell Telephone Laboratories, who created an environment that both attracted the very best researchers and made space for their creativity. The Nobel Prize in Physics 1956 was awarded jointly to William Shockley, John Bardeen and Walter Brattain "for their researches [sic] on semiconductors and their discovery of the transistor effect" at "Bell Labs" (then "Bell Telephone Laboratories", now "Nokia Bell Labs"), precisely fifty years after Thomson's prize for the discovery of the electron.

So now we live in the "Silicon Age", with silicon transistors forming the core of much of the microelectronics that enable our modern way of life. Not only our computers and mobile phones, but every aspect of for example commerce, transportation, and communication are now underpinned by microelectronic devices. Since those very first transistors in the 1940s and 50s, we have improved the properties of silicon devices to an astonishing extent, enabling the transformation for example from clunky old main-frame computers to sleek smartphones with tremendous capabilities, all with the same material – silicon – at their core. We have grown to expect the exponential increase in capability and corresponding decrease in size and cost, captured by Moore's Law [2], and to anticipate ever more automation and convenience in our everyday activities.

But this silicon revolution will soon be forced to come to an end as we start to run into fundamental physical limits, set by the size of the individual atoms that make up the silicon material. And this means that the steady march towards faster, smaller lighter products with more and more functionality can't continue within our existing framework. Now, while this might not seem so disastrous (certainly the controls on my smartphone are already smaller than I can see without my reading glasses), it is in fact a profound problem for society: As living standards improve in emerging regions and the "internet of things" becomes more widespread, worldwide use of microelectronics is expanding more rapidly than ever before, so that by most projections more than half of the world's energy will be consumed by information technologies within a couple of decades [3]. And this is not sustainable. So, we need to take the step beyond the silicon age, we need to develop an entirely new device paradigm, and to do this we need a new material. Without a new material, we are stuck with our existing concepts for information technology and we have an energy bottleneck in human progress. And fundamental research in Materials Science – very likely with a complete change in direction – underpins the invention of this material.

Let me give you an example from my own research. A couple of decades ago, I was a young postdoctoral researcher working on ferromagnetic materials – these are materials that contain magnetic dipoles with their north and south poles aligned parallel to each other – in a research group that specialized in ferroelectric materials, which are materials that contain aligned electric dipoles, made of positive and negative charges. My plan was to take the tools and techniques that my host group had developed to study ferroelectric materials and apply them to the study of ferromagnetic materials; the "ferro" in both names reflects the similarities in some of the underlying physics between the two material classes.

I noticed, almost by accident, that the kinds of materials that I was working on were different in many ways from those of my colleagues. For example, most ferromagnetic materials are black metals, like iron, whereas most ferroelectric materials are transparent oxide ceramics; barium titanate, chemical formula $BaTiO_3$, is the prototypical example. My materials were shiny and ductile, theirs brittle. This apparent "contra-indication" between ferromagnetism and ferroelectricity intrigued me, and after a weekend of poring over encyclopediae of both material types (this was before the days of convenient on-line searching!) I convinced myself that it was real: There were no ferromagnetic materials in the handbook on ferroelectrics [4], and vice versa. Immediately I asked myself the question "Why are there so few magnetic ferroelectrics?" [5]. Answering this question became a passion (maybe even an obsession) for me and formed the focal point of my research program over the next decades. I changed direction, and stopped heading where I was headed.

Finding the answer took fundamental research into the basic chemistry of the bonding in ferroelectric materials to understand why it contra-indicated ferromagnetism. And this fundamental research allowed us to make what was in the end quite a simple discovery: That the atoms that form the kinds of chemical bonds needed to produce electric dipoles in a material have different arrangements of their constituent electrons from those that tend to make magnetic dipoles. But we were also able to show that there is no fundamental law of physics preventing their coexistence. Armed with this understanding of why ferromagnetic and ferroelectrics tend not to occur together, my colleagues and I were able to create new materials – we call them multiferroics – that really are ferromagnetic and ferroelectric. We did this in two ways: Our first route was to design new materials that combine the two types of atoms – those that tend to form magnetic dipoles and those that tend to form electric dipoles – in the same material. An example is the perovskite-structure oxide, bismuth ferrite, $BiFeO_3$, in which the iron atoms provide the magnetism, and the ferroelectricity comes from the so-called "lone pair" of electrons on the Bi atoms [6]. (Readers with a chemistry background will recognize an analogy here with the origin of the dipole moment in the ammonia molecule.) Our second method was to engineer new crystal structures that force magnetic atoms into new environments that are compatible with electric dipoles, which is the case for example with yttrium manganite, $YMnO_3$ [7]. The unconventional mechanism for ferroelectricity in the latter case causes an unusual arrangement of the orientation regions of the electric dipoles (called domains) resulting in exquisite textures, like those shown in the header to this article for the isomorphous $ErMnO_3$.

I emphasize here that this research was driven entirely by annoyance that such a simple question – Why are there so few magnetic ferroelectrics? – had not been answered [8]. At the start of our work, there were no device physicists waiting eagerly for our materials, because no-one was thinking about the possibilities that a material that is both magnetic and ferroelectric might offer. Practical, working multiferroics did not exist even in our imaginations. Soon, however, we discovered that these aesthetic crystal chemistries, with their gorgeous dipolar domain structures and their combined magnetism and ferroelectricity, have entirely unexpected and potentially technologically transformative functionalities [9]. Perhaps most importantly, we demonstrated that we are able to modify the magnetic properties of multiferroics with electric fields [10, 11]. This is exciting from a basic physics perspective – usually a magnetic field is needed to modify magnetic properties – but also has profound technological implications: Replacing the magnetic fields in our existing magnetism-based technologies with electric fields offers tremendous opportunity for energy savings, miniaturization and efficiency. In a completely unexpected discovery, we found that the domain walls – the intersections separating regions (domains) with different orientations of the ferroelectric dipoles – form nanoscale conducting channels that can be moved around using electric fields [12]. This has potential application in novel memory or information processing architectures. The combination of magnetism and ferroelectricity leads to an unusual surface electronic structure that is being actively explored for catalysis and water splitting applications. And the ability to control the electrical and structural properties using magnetic fields, which can be applied without invasive electrodes and wires, is being explored for biomedical applications. Our new multiferroic materials, which started out as a playground for exploring fundamental questions in physics and chemistry, are poised to enable new device paradigms, and in turn entirely new ways of designing technologies [13]. Perhaps we are about to enter a new *Multiferroics Age*?

Or perhaps not. Of course, there's more to human civilization than information technology, and more to Materials Science than microelectronics. And we need fundamental research in all branches of Materials Science to address many of the most challenging global issues identified by the United Nations [14]. Problems of climate change, and the environment, for example, will only be solved with new materials that can provide clean affordable alternative energy. Improved bio- and bio-compatible materials are needed to advance human health and to assist persons with physical disabilities. New materials made from earth-abundant, readily available elements will ensure a more equitable global wealth distribution and mitigate our dependency on minerals mined in conflict zones. My personal hope is that historians will consider the post-silicon era to be a "Golden Age" in which fundamental research in Materials Science will have helped to enable a world in which peace, prosperity and reason prevail.

So, what next? Well, like many others in the Materials Physics community, I'm working to understand the so-called strong correlations between electrons in solids. Why, if one electron somewhere in a material rearranges a little bit, this explicitly and profoundly affects all of the other electrons. This research is very fundamental and might never lead to anything useful. Even in that case I would argue that it is worthwhile: Exposing the profound beauty of interacting electrons is comparable to imaging the complexity of our galaxy, the satisfaction of finding a new elementary particle at CERN, or the joy of listening to the Tonhalle Orchestra play a Brahms symphony; all activities which as a society we find worthwhile to invest in. On the other hand, understanding strong electron correlations could be the first step towards making a room-temperature superconductor, a material that conducts electricity without any resistance, under everyday conditions.

Such a material would revolutionize energy production, transmission and storage: Imagine power grids that don't lose energy, portable MRI machines, cheap and widespread "Maglev" trains and paradigm shifts in computing technologies. A room-temperature superconductor would be utterly geopolitically transformative. Then I would bet that the next era of human civilization would be named after this as-yet undiscovered material.

Let me end with a plea to government officials, managers of funding agencies, and university administrators: Of course, applied research is important, nowhere more so than in Materials Science. And every Materials Scientist is in her heart an engineer, strongly motivated by solving practical problems that will enable the technologies that improve people's lives. We spend most of our time setting practical achievable goals for relevant problems and developing materials that get us to where we are headed. But if we work only on materials with an application already in mind we limit ourselves to applications that we have already thought of. And we will not make something really new that will open up entirely new directions and device paradigms. The true breakthroughs that will change the course of history will not come from initiatives to improve existing materials or devices, or to advance technologies that have already been identified. Instead, they will come from off-beat individuals or small teams of fundamental researchers pushing the boundaries of knowledge in directions for which there is not yet an application. Pioneers who will not end up where the rest of us are headed, but instead will change direction and go somewhere that we have not yet envisaged. I urge you to create an environment that nurtures that adventurous spirit, an environment that enables not only the applied research that will benefit society immediately, but also the fundamental research that we can enjoy now for its aesthetic beauty and that will have its technology payoff only in future generations. I urge you to be good scientific ancestors. And what better legacy than to have enabled the discovery of the material that defines the course of human civilization. ∎

### References

[1] J. J. Thomson, *Cathode rays*, Philosophical Magazine, Series **5**, Volume **44**, No. **269**, 293 (1897).

[2] G. E. Moore, *Cramming more components onto integrated circuits*, Electronics, **38**, 114 (1965).

[3] *www.iea.org/publications/freepublications/publication/gigawatts2009.pdf*

[4] M. E. Lines and A. M. Glass, *Principles and applications of ferroelectrics and related materials*, Oxford University Press 2001 (first published 1977).

[5] N. A. Hill (now Spaldin), *Why are there so few magnetic ferroelectrics?*, J. Phys. Chem. B **104**, 6694 (2000).

[6] J. Wang, J. B. Neaton, H. Zheng, V. Nagarajan, S. B. Ogale, B. Liu, D. Viehland, V. Vaithyanathan, D. G. Schlom, U. V. Waghmare, N. A. Spaldin, K. M. Rabe, M. Wuttig and R. Ramesh, *Epitaxial BiFeO$_3$ multiferroic thin film heterostructures*, Science **299**, 1719 (2003).

[7] B. B. van Aken, T. T. M. Palstra, A. Filippetti and N. A. Spaldin, *The origin of ferroelectricity in magnetoelectric YMnO₃*, Nature Materials **3**, 164 (2004).

[8] N. A. Spaldin, *Find your most interesting question*, Science **349**, 110 (2015).

[9] N. A. Spaldin and M. Fiebig, *The renaissance of magnetoelectric multiferroics*, Science **309**, 391 (2005).

[10] T. Zhao, A. Scholl, F. Zavaliche, K. Lee, M. Barry, A. Doran, M. P. Cruz, Y. H. Chu, C. Ederer, N. A. Spaldin, R. R. Das, D. M. Kim, S. H. Baek, C. B. Eom and R. Ramesh, *Electrical control of antiferromagnetic domains in multiferroic BiFeO3 films at room temperature*, Nature Materials **5**, 823 (2006).

[11] N. A. Spaldin and R. Ramesh, *Electric field control of magnetism in complex oxide thin films,* Materials Research Society Bulletin **33**, 1047 (2008).

[12] J. Seidel, L. W. Martin, Q. He, Q. Zhan, Y.-H. Chu, A. Rother, M. E. Hawkridge, P. Maksymovych, P. Yu, M. Gajek, N. Balke, S. V. Kalinin, S. Gemming, F. Wang, G. Catalan, J. F. Scott, N. A. Spaldin, J. Orenstein and R. Ramesh, *Conduction at domain walls in oxide multiferroics*, Nature Materials **8**, 229 (2009).

[13] N. A. Spaldin, *Multiferroics: Past, present, and future*, Materials Research Society Bulletin **42**, 385 (2017).

[14] *http://www.un.org/en/sections/issues-depth/global-issues-overview/*